\RequirePackage{ifpdf}
\ifpdf 
\documentclass[pdftex]{sigma}
\else
\documentclass{sigma}
\fi

\numberwithin{equation}{section}

\numberwithin{remark}{section}

\begin{document}

\allowdisplaybreaks

\renewcommand{\thefootnote}{$\star$}

\renewcommand{\PaperNumber}{043}

\FirstPageHeading

\ShortArticleName{Recursions of Symmetry Orbits and Reduction without Reduction}

\ArticleName{Recursions of Symmetry Orbits and Reduction\\ without Reduction\footnote{This paper is a
contribution to the Special Issue ``Symmetry, Separation, Super-integrability and Special Functions~(S$^4$)''. The
full collection is available at
\href{http://www.emis.de/journals/SIGMA/S4.html}{http://www.emis.de/journals/SIGMA/S4.html}}}

\Author{Andrei A.~MALYKH~$^\dag$ and Mikhail B.~SHEFTEL~$^\ddag$}

\AuthorNameForHeading{A.A.~Malykh and M.B.~Sheftel}

\Address{$^\dag$~Department of Numerical Modelling, Russian State
Hydrometeorlogical University,\\
\hphantom{$^\dag$}~Malookhtinsky
pr. 98, 195196 St. Petersburg, Russia}
\EmailD{\href{mailto:andrei-malykh@mail.ru}{andrei-malykh@mail.ru}}

\Address{$^\ddag$~Department of Physics, Bo\u{g}azi\c{c}i
University 34342 Bebek, Istanbul, Turkey}
\EmailD{\href{mailto:mikhail.sheftel@boun.edu.tr}{mikhail.sheftel@boun.edu.tr}}

\ArticleDates{Received January 29, 2011, in f\/inal form April 25, 2011;  Published online April 29, 2011}

\Abstract{We consider a four-dimensional PDE possessing partner symmetries
mainly on the example of complex Monge-Amp\`ere equation (CMA).
We use simultaneously two pairs of symmetries related by a recursion relation,
which are mutually complex conjugate for CMA.  For
both pairs of partner symmetries, using Lie equations, we introduce
explicitly group parameters as additional variables, replacing
symmetry characteristics and their complex conjugates by
derivatives of the unknown with respect to group parameters. We study
the resulting system of six equations in the eight-dimensional
space, that includes CMA, four equations of the recursion
between partner symmetries and one integrability condition of this
system. We use point symmetries of this extended system for
performing its symmet\-ry reduction with respect to group parameters
that facilitates solving the extended system. This procedure does
not imply a reduction in the number of physical variables and
hence we end up with orbits of non-invariant solutions of
CMA, generated by one partner symmetry, not used in the
reduction. These solutions are determined by six linear equations
with constant coef\/f\/icients in the f\/ive-dimensional space which are obtained
by a three-dimensional Le\-gendre transformation of the reduced extended system. We present
algebraic and exponential examples of such solutions that govern Legendre-transformed Ricci-f\/lat
K\"ahler metrics with no Killing vectors. A similar procedure is brief\/ly outlined for Husain equation.}


\Classification{35Q75; 83C15}

\section{Introduction}

In our recent papers \cite{shma,mns,mnsh,lift,shma08} we
demonstrated how to use partner symmetries for ob\-tai\-ning
non-invariant solutions of four-dimensional PDEs of Monge-Amp\`ere type,
which determine Ricci-f\/lat metrics with no Killing vectors in the anti-self-dual gravity.
Here we present another method of solving this problem by introducing explicitly group parameters of
two pairs of partner symmetries (mutually complex conjugate for real equations like CMA) as additional
variables in the system of four equations that determine
recursions between the partner symmetries. For this purpose, we use Lie
equations of partner symmetries for replacing their symmetry
characteristics with derivatives of the unknown with respect
to group parameters. Such an approach can be applied to any non-linear PDE of the Monge-Amp\`ere type
which admits a symmetry condition to be of a divergence form.
The resulting system of f\/ive PDEs, that also includes the studied equation of the Monge-Amp\`ere type,
has one non-trivial integrability condition,
which means that both pairs of partner symmetries commute with each other,
so that we end up with six PDEs for one unknown in the eight-dimensional space of independent variables.

The problem of f\/inding a solution of this extended system becomes
much easier if we use three point symmetries of this system for
performing its symmetry reduction with respect to a combination of
variables involving group parameters. (One symmetry is not used
for the reduction and so we still keep the dependence of solutions
on one group parameter.) This does not mean a~reduction in the
number of ``physical variables'' because only group parameters
disappear from the set of independent variables. Hence we still
obtain a noninvariant solution that depends on the four physical
variables and one remaining group parameter. Performing Legendre
transformation in two physical variables and one group
parameter we arrive at a~system of six linear equations with
constant coef\/f\/icients for one unknown with f\/ive independent
variables. One can easily construct many explicit solutions of
this system which govern the Legendre-transformed heavenly metric.
To illustrate this program, we consider CMA as a main example
together with a brief outline of a similar procedure for the
Husain equation. For the elliptic CMA we construct explicitly
algebraic and exponential solutions.

We emphasize that our task here is not to obtain
solutions of the original CMA equation, which would require
application of the inverse Legendre transformation in two variables to our solutions,
but rather to construct Ricci-f\/lat vacuum metrics (solutions
of Einstein equations with either Euclidean or neutral signature) with no Killing vectors
(continuous symmetries).
For this purpose, we apply the inverse one-dimensional Legendre transformation
with respect to the group parameter to noninvariant solutions of the linear system
which results in solutions of the Legendre-transformed CMA that determine metrics
with the above properties. The general form of these metrics is obtained by a two-dimensional
Legendre transformation of the K\"ahler metric with respect to a couple of complex conjugate physical variables.

In Section \ref{sec-eqs}, we derive an over-determined
system of six equations in eight-dimensional space with one
unknown and a set of eight independent variables containing
four symmetry group parameters. This system does not generate further
independent second-order integrability conditions and it determines a class
of solutions of the elliptic and hyperbolic CMA.

In Section \ref{sec-symred}, we list all point symmetries of this
extended system and perform a symmetry reduction with respect to
two group parameters. We calculate all point symmetries of the
resulting system of six equations in six-dimensional space and use
one of the symmetries for further reduction of this system with
respect to one more group parameter, with one group parameter
still remaining as the f\/ifth independent variable.

In Section \ref{sec-Legendre} we show how by applying Legendre transformation
in three variables, including the group parameter, one arrives at a system of linear
partial dif\/ferential equations with constant coef\/f\/icients for one unknown
in a f\/ive-dimensional space, which does not generate further independent
second-order dif\/ferential constraints. Many explicit exact solutions of this
system can easily be obtained.

In Section \ref{sec-kahler}, we obtain the Legendre transform of the
K\"ahler heavenly metric, using Legendre transformation with
respect to two ``physical'' variables only. One should be careful
to perform correctly three-dimensional Legendre transformation of the metric
involving a transformation with respect to the group parameter. Otherwise,
the dif\/ferential of the group parameter appears in the transformed metric
which results in the metric not being Ricci-f\/lat anymore. Therefore, in the
next section we have to apply
the inverse one-dimensional Legendre transformation in the group parameter
to solutions of the linear system in order to obtain the potential $v$, that
governs the Legendre-transformed metric.

In Section \ref{sec-solutions}, we present examples of exact algebraic and
exponential non-invariant solutions
of our equations which determine Ricci-f\/lat metrics not admitting
Killing vectors.

In Section \ref{sec-husain}, we derive an overdetermined system of six PDEs
with one unknown and a~set of eight variables containing four symmetry group parameters,
which determines a class of solutions of the Husain equation. This system includes one
second-order integrability condition of the other f\/ive equations and it
has no further second-order integrability conditions.

\section{Basic equations}
 \label{sec-eqs}

We are interested in non-invariant solutions of the
complex Monge-Amp\`ere equation (CMA)
\begin{equation}
  u_{1\bar 1}u_{2\bar 2} - u_{1\bar 2}u_{2\bar 1} = \varepsilon ,
  \qquad \varepsilon = \pm 1 ,\label{cma}
\end{equation}
which is closely related to the f\/irst heavenly equation of
Pleba\~nski \cite{pleb}. Symmetry condition, that determines
symmetry characteristics \cite{olv} $\varphi$ of \eqref{cma},
\begin{equation}
  u_{1\bar 1}\varphi_{2\bar 2} + u_{2\bar 2}\varphi_{1\bar 1}
 - u_{1\bar 2}\varphi_{2\bar 1} - u_{2\bar 1}\varphi_{1\bar 2} = 0 \label{symcond}
\end{equation}
can be set in the total divergence form
\[
  (u_{1\bar 1}\varphi_2  - u_{2\bar 1}\varphi_1)_{\bar 2}
  - (u_{1\bar 2}\varphi_2 - u_{2\bar 2}\varphi_1)_{\bar 1} = 0.
\]
This suggests local existence of potential $\psi$ def\/ined by the
equations
\begin{equation}
 \psi_{\bar 1} = u_{1\bar 1}\varphi_2  - u_{2\bar 1}\varphi_1,
 \qquad \psi_{\bar 2} = u_{1\bar 2}\varphi_2 - u_{2\bar 2}\varphi_1 \label{12}
\end{equation}
together with the complex conjugate equations
\begin{equation}
 \bar\psi_1 = u_{1\bar 1}\bar\varphi_{\bar 2} - u_{1\bar 2}\bar\varphi_{\bar
 1}, \qquad \bar\psi_2 = u_{2\bar 1}\bar\varphi_{\bar 2} - u_{2\bar 2} \bar\varphi_{\bar 1} . \label{b12}
\end{equation}
A straightforward check shows that the potential $\psi$ also satisf\/ies symmetry condition
\eqref{symcond}, so that $\psi$ is also a symmetry if $\varphi$ is a symmetry and hence the relations \eqref{12}
and~\eqref{b12} are recursion relations for symmetries (``partner symmetries'').
Transformation~\eqref{12} is algebraically invertible since
its determinant is equal to~$\varepsilon$ due to~\eqref{cma}. Inverse
transformation has the form
\begin{equation}
 \varphi_1 = \varepsilon(u_{1\bar 2}\psi_{\bar 1}  - u_{1\bar 1}\psi_{\bar 2}),
 \qquad \varphi_2 = \varepsilon(u_{2\bar 2}\psi_{\bar 1} - u_{2\bar
 1}\psi_{\bar 2}) \label{34}
\end{equation}
together with its complex conjugate
\begin{equation}
 \bar\varphi_{\bar 1} = \varepsilon(u_{2\bar 1}\bar\psi_1 - u_{1\bar
 1}\bar\psi_2), \qquad \bar\varphi_{\bar 2} = \varepsilon(u_{2\bar 2}\bar\psi_1
 - u_{1\bar 2} \bar\psi_2) .  \label{b34}
\end{equation}
For symmetries with characteristics $\varphi$, $\bar\varphi$,
$\psi$ and $\bar\psi$, Lie equations read
\begin{equation}
\varphi = u_\tau,\qquad \bar\varphi = u_{\bar\tau}, \qquad \psi =
u_\sigma,\qquad \bar\psi = u_{\bar\sigma} ,\label{Lie}
\end{equation}
where $\tau$, $\sigma$ together with their complex conjugates
$\bar\tau$, $\bar\sigma$ are group parameters. Simultaneous inclusion
of several group parameters as additional independent variables
implies the commutativity conditions for corresponding symmetries in the form
\[\varphi_{\bar\tau} = \bar\varphi_\tau,\qquad\psi_{\bar\sigma} = \bar\psi_\sigma,\qquad
    \varphi_\sigma = \psi_\tau,\qquad \varphi_{\bar\sigma} = \bar\psi_\tau \]
and complex conjugates to the last two equations.

We now use~\eqref{Lie} to replace symmetry characteristics by derivatives of
$u$ with respect to group parameters in equations~\eqref{12},
\eqref{34} and their complex conjugates~\eqref{b12},~\eqref{b34}
with the result
\begin{alignat}{3}
& u_{\sigma\bar 1} = u_{1\bar 1}u_{\tau 2} - u_{2\bar 1}u_{\tau 1},
\qquad && u_{\sigma\bar 2} = u_{1\bar 2}u_{\tau 2} - u_{2\bar
2}u_{\tau 1}, & \label{12a} \\
& u_{\tau 1} = \varepsilon(u_{1\bar 2}u_{\sigma\bar 1} - u_{1\bar 1}u_{\sigma\bar
 2}), \qquad && u_{\tau 2} = \varepsilon(u_{2\bar 2}u_{\sigma\bar 1}
 - u_{2\bar 1}u_{\sigma\bar 2}), & \label{34a}
\end{alignat}
and the complex conjugate equations
\begin{gather}
 u_{\bar\sigma 1} = u_{1\bar 1}u_{\bar\tau\bar 2} - u_{\bar 21}u_{\bar\tau\bar 1},
 \qquad u_{\bar\sigma 2} = u_{\bar 12}u_{\bar\tau\bar 2}
 - u_{2\bar 2}u_{\bar\tau\bar 1}, \label{b12a} \\
 u_{\bar\tau\bar 1} = \varepsilon(u_{\bar 12}u_{\bar\sigma 1} - u_{1\bar 1}u_{\bar\sigma 2}),
 \qquad u_{\bar\tau\bar 2} = \varepsilon(u_{2\bar 2}u_{\bar\sigma 1} - u_{\bar 21}u_{\bar\sigma 2}).
  \label{b34a}
\end{gather}
We note that four equations \eqref{34a} and \eqref{b34a} are
algebraic consequences of other equa\-tions~\eqref{12a},
\eqref{b12a} and CMA. We note also that CMA itself follows as
an algebraic consequence from equations~\eqref{12a} and~\eqref{b12a}. We choose equations
\eqref{12a}, \eqref{b12a} and the f\/irst equation in~\eqref{b34a} (since we regard CMA as algebraically dependent equation)
as a set of f\/ive algebraically independent equations.

 One can easily check that cross-dif\/ferentiations of these equations
with respect to group parameters yield only identities, e.g. $(u_{\sigma\bar 1})_{\bar\tau} = (u_{\bar\tau\bar 1})_\sigma$ implies
\[(u_{1\bar 1}u_{\tau 2} - u_{2\bar 1}u_{\tau 1})_{\bar\tau} = \varepsilon(u_{2\bar 1}u_{1\bar\sigma} - u_{1\bar 1}u_{2\bar\sigma})_\sigma ,\]
which is identically satisf\/ied due to existing equations.

To study further dif\/ferential integrability conditions of our system, we
set the f\/irst equations in \eqref{12a} and \eqref{b34a} in the form
\begin{equation}
  (u_{1\bar 1}u_2)_\tau = (u_\sigma + u_2u_{\tau 1})_{\bar
  1},\qquad (u_{1\bar 1}u_2)_{\bar\sigma} =(u_2u_{\bar\sigma 1} - \varepsilon
  u_{\bar\tau})_{\bar 1}. \label{active}
\end{equation}
Equations \eqref{active} constitute an active system since they
have a second order nontrivial integrability condition obtained by
cross-dif\/ferentiation of these equations with respect to
$\bar\sigma$ and $\tau$ and further integration with respect to
${\bar z}^1$
\[
 \varepsilon u_{\tau\bar\tau} + u_{\sigma\bar\sigma} + u_{\bar\sigma 2}u_{\tau
 1} - u_{\bar\sigma 1}u_{\tau 2} = 0,
\]
where the ``constant'' of integration can be eliminated by a redef\/inition of $u$. To make
this equation self-conjugate, we multiply it with an overall
factor $u_{1\bar 1}$ and then eliminate $u_{1\bar 1}u_{\bar\sigma
2}$ and $u_{1\bar 1}u_{\tau 2}$ in the last two terms of
\eqref{active} using f\/irst equations in \eqref{b34a} and
\eqref{12a}, respectively, with the f\/inal form of the
integrability condition
\begin{equation}
u_{1\bar 1}(\varepsilon u_{\tau\bar\tau} + u_{\sigma\bar\sigma}) - \varepsilon u_{\tau 1}
u_{\bar\tau\bar 1} - u_{\sigma\bar 1}u_{\bar\sigma 1} = 0. \label{sixeq}
\end{equation}
In a similar way we obtain another form of the integrability
condition
\[
u_{2\bar 2}(\varepsilon u_{\tau\bar\tau} + u_{\sigma\bar\sigma}) - \varepsilon u_{\tau 2}
u_{\bar\tau\bar 2} - u_{\sigma\bar 2}u_{\bar\sigma 2} = 0.
\]
We can choose \eqref{12a}, \eqref{b12a}, \eqref{sixeq} and f\/irst
equation in \eqref{b34a} for the set of algebraically independent
equations. All other equations, including CMA, are linearly
dependent on the chosen equations. One could also check that there
are no further independent second-order dif\/ferential integrability conditions
of our system of six equations.

\section{Symmetry reduction of extended system}
 \label{sec-symred}

From now on we set $\varepsilon = + 1$ and thereby concentrate on the elliptic CMA as an example.
We list the generators of all point symmetries of the extended
system of six equations CMA, \eqref{12a}, \eqref{b12a} and
\eqref{sixeq} (f\/irst equation in \eqref{b34a} also admits these
symmetries as an obvious consequence)
\begin{gather}
 X_1 = \partial_\tau ,\quad \bar X_1 = \partial_{\bar\tau},\qquad X_2 = \partial_\sigma, \qquad \bar X_2 = \partial_{\bar\sigma} ,
 \qquad X_3 = \tau\partial_\tau + \sigma\partial_\sigma, \nonumber \\
 \bar X_3 = \bar\tau\partial_{\bar\tau} + \bar\sigma\partial_{\bar\sigma},\qquad
 X_4 = z^2\partial_2 - \bar{z}^2\partial_{\bar 2} +
\bar\tau\partial_{\bar\tau} - \tau\partial_\tau +
\sigma\partial_\sigma - \bar\sigma\partial_{\bar\sigma} ,
\nonumber \\
 X_5 = \tau\partial_{\bar\sigma} -
 \sigma\partial_{\bar\tau}, \qquad \bar X_5 = \bar\tau\partial_\sigma -
 \bar\sigma\partial_\tau,
\qquad X_6 = z^2\partial_2 + \bar{z}^2\partial_{\bar 2} +
u\partial_u , \nonumber \\
 X_a = a(z^1,z^2,\bar\tau,\sigma)\partial_u ,\qquad
 X_{\bar a} = \bar a(\bar{z}^1,\bar{z}^2,\tau,\bar\sigma)\partial_u ,\label{symgen} \\
 X_c = c_{z^1}\partial_2 - c_{z^2}\partial_1 + (\tau c_\sigma - \bar\sigma c_{\bar\tau})\partial_u,
 \nonumber \\
 X_{\bar c} = \bar c_{\bar{z}^1}\partial_{\bar 2} - \bar c_{\bar{z}^2}\partial_{\bar 1}
 + (\bar\tau \bar c_{\bar\sigma} - \sigma \bar c_\tau)\partial_u,\qquad X_f =
 f(\tau,\sigma,\bar\tau,\bar\sigma)\partial_u ,
 \nonumber
\end{gather}
where $a$, $\bar a$, $c=c(z^1,z^2,\bar\tau,\sigma)$,
$\bar c = \bar c(\bar{z}^1,\bar{z}^2,\tau,\bar\sigma)$ are arbitrary functions and
$f(\tau,\sigma,\bar\tau,\bar\sigma)$ satisf\/ies the equation
$f_{\tau\bar\tau} + f_{\sigma\bar\sigma} = 0$. We note that
obvious translational symmetry generators $\partial_1$,
$\partial_{\bar 1}$, $\partial_2$ and $\partial_{\bar 2}$ are
particular cases of the generators $X_c$ and $X_{\bar c}$.

We also note that if we perform the symmetry reduction with
respect to the two symmetries $X_2 - \gamma X_1$ and $\bar X_2 - \bar\gamma
\bar X_1$ in~\eqref{symgen} simultaneously, then solutions satisfy
the additional constraints $u_\sigma = \gamma u_\tau$ and
$u_{\bar\sigma} = \bar\gamma u_{\bar\tau}$. In this case our
equation~\eqref{sixeq} becomes
\begin{equation}
  (1 + \gamma\bar\gamma)u_{1\bar 1}u_{\tau\bar\tau} -
  u_{1\tau}u_{\bar 1\bar\tau} - \gamma \bar\gamma u_{1\bar\tau}
  u_{\bar 1\tau} = 0, \label{fer}
\end{equation}
that is equivalent to the general heavenly equation of B.~Doubrov
and E.~Ferapontov~\cite{ferap}, which depends essentially on a
single parameter.

We specify two symmetries from \eqref{symgen} for a symmetry
reduction of the extended system
\begin{equation}
 X_I =   \partial_\tau - \lambda\partial_1, \qquad
\bar X_I = \partial_{\bar\tau} - \bar\lambda\partial_{\bar 1},
  \label{sym2}
\end{equation}
where $\lambda$ is a constant parameter with $|\lambda|=1$, so
that $\bar\lambda = \lambda^{-1}$. Solutions of CMA invariant
with respect to symmetries \eqref{sym2} are determined by the
conditions
\begin{equation}
  u_\tau = \lambda u_1,\qquad u_{\bar\tau} = \lambda^{-1} u_{\bar
  1}. \label{invcond}
\end{equation}
Using \eqref{invcond}, we eliminate $u_\tau$ and $u_{\bar\tau}$ in
all the equations \eqref{12a}, \eqref{b12a}, f\/irst equation in
\eqref{b34a} and \eqref{sixeq} to obtain
\begin{gather}
 u_{\sigma\bar 1} = \lambda(u_{1\bar 1}u_{12} - u_{2\bar 1}
 u_{11}),\qquad
 u_{\sigma\bar 2} = \lambda(u_{1\bar 2}u_{12} - u_{2\bar 2}u_{11}),
 \label{I_II} \\
 u_{\bar\sigma 1} = \lambda^{-1}(u_{1\bar 1}u_{\bar 1\bar 2} - u_{\bar 21}u_{\bar 1\bar 1}),
 \qquad u_{\bar\sigma 2} = \lambda^{-1}(u_{\bar 12}u_{\bar 1\bar 2} -
u_{2\bar 2}u_{\bar 1\bar 1}),
 \label{bI_II} \\
 u_{\bar 1\bar 1} = \lambda(u_{\bar 12}u_{\bar\sigma 1} - u_{1\bar 1}u_{\bar\sigma
 2}), \label{bIII} \\
 u_{1\bar 1}u_{\sigma\bar\sigma} - u_{1\bar\sigma}u_{\bar
 1\sigma} = u_{11}u_{\bar 1\bar 1} - u_{1\bar 1}^2 .  \label{8}
\end{gather}
We note that equation \eqref{8} can be obtained by the Legendre
transformation
\[
  v = u - z^1u_1 - \bar{z}^1u_{\bar 1},\qquad p = - u_1,\qquad \bar p = - u_{\bar 1}
\]
of the CMA in new variables
\[
v_{p\bar p}v_{\sigma\bar\sigma} - v_{p\bar\sigma}v_{\sigma\bar p}
= 1 .
\]

All point symmetry generators of the system of equations CMA,
\eqref{I_II}, \eqref{bI_II} and \eqref{8} (and the equation in
\eqref{bIII} as a consequence) are listed below
\begin{gather}
 X_1 = z^1\partial_1 - \bar{z}^1\partial_{\bar 1} -
2(z^2\partial_2 - \bar{z}^2\partial_{\bar 2}),\qquad X_2 =
z^2\partial_2 + \bar{z}^2\partial_{\bar 2} + u\partial_u , \nonumber \\
 X_3 = a(\sigma)\partial_2 +
\frac{1}{2\lambda}(z^1)^2a'(\sigma)\partial_u ,\qquad X_4 =
b(\bar\sigma)\partial_{\bar 2} + \frac{\lambda}{2}(\bar{z}^1)^2
b'(\bar\sigma)\partial_u, \nonumber \\
 X_5 = c'(\sigma)(z^1\partial_1 - z^2\partial_2) +
 c(\sigma)\partial_\sigma -
 \frac{1}{2\lambda}(z^1)^2z^2c''(\sigma)\partial_u, \nonumber \\
 X_6 = d'(\bar\sigma)(\bar{z}^1\partial_{\bar 1} - \bar{z}^2\partial_{\bar 2}) +
 d(\bar\sigma)\partial_{\bar\sigma} -
 \frac{\lambda}{2}(\bar{z}^1)^2\bar{z}^2d''(\bar\sigma)\partial_u,
 \nonumber \\
 X_7 = - \lambda f_{z^2}(z^2,\sigma)\partial_1 +
 z^1f_\sigma(z^2,\sigma)\partial_u, \qquad
 X_8 = - \frac{1}{\lambda}
 g_{\bar{z}^2}(\bar{z}^2,\bar\sigma)\partial_{\bar 1} + \bar{z}^1
 g_{\bar\sigma}(\bar{z}^2,\bar\sigma)\partial_u ,\nonumber \\
 X_9 = h(z^2,\sigma)\partial_u,\qquad X_{10} =
 k(\bar{z}^2,\bar\sigma)\partial_u .
 \label{pointsym}
\end{gather}

To simplify the problem of solving our equations, we make another
symmetry reduction with respect to group parameters that does not
imply a reduction in the original ``physical'' variables. For this
purpose, we use the symmetry $X_5 - iX_6 = i (\partial_\sigma -
\partial_{\bar\sigma})$ from the list \eqref{pointsym} taken at
$c(\sigma)=1$ and $d(\bar\sigma)=1$. The invariance condition for
solutions is $u_\sigma = u_{\bar\sigma} = u_\rho$, where $\rho =
\sigma + \bar\sigma$ is the invariant variable which replaces
$\sigma$ and $\bar\sigma$ as an argument of $u$. Equations
\eqref{I_II} and \eqref{bI_II} become
\begin{equation}
u_{\rho\bar 1} = \lambda(u_{1\bar 1}u_{12} - u_{2\bar 1}
u_{11}),\qquad  u_{\rho\bar 2} = \lambda(u_{1\bar 2}u_{12}
- u_{2\bar 2}u_{11}) \label{redI_II}
\end{equation}
together with their complex conjugates, while f\/irst equation in
\eqref{b34a} becomes
\begin{equation}
u_{\bar 1\bar 1} = \lambda(u_{\bar 12}u_{\rho 1} - u_{1\bar
1}u_{\rho 2}) \label{bIIIred}
\end{equation}
 and equation \eqref{8} takes the form
\begin{equation}
u_{1\bar 1}u_{\rho\rho} - u_{1\rho}u_{\bar
 1\rho} = u_{11}u_{\bar 1\bar 1} - u_{1\bar 1}^2 .
 \label{red8}
\end{equation}

\section{Linearizing Legendre transformation}
 \label{sec-Legendre}

Equation \eqref{red8} is linearized by the three-dimensional
Legendre transformation
\begin{equation}
  u = w - pw_p - \bar pw_{\bar p} - rw_r,\qquad z^1 = - w_p,\qquad
  \bar{z}^1 = - w_{\bar p},\qquad \rho = - w_r ,\label{Leg3}
\end{equation}
where $w = w(p,\bar p,r,z,\bar{z})$ depends on the new variables
$p$, $\bar p$, $r$, $z=z^2$, $\bar{z}=\bar{z}^2$. Equation \eqref{red8}
takes the form $(w_{p\bar p} + w_{rr})/D  = 0$,
where
\begin{equation}
D \stackrel{def}{=} \left|
\begin{array}{ccc}
 w_{pp} & w_{p\bar p} & w_{pr} \\
 w_{\bar pp} & w_{\bar p\bar p} & w_{\bar pr} \\
 w_{rp} & w_{r\bar p} & w_{rr}
\end{array}
\right| \ne 0  \label{det}
\end{equation}
is the existence condition for Legendre transformation
\eqref{Leg3}. Thus, equation \eqref{red8} becomes a~linear equation
\begin{equation}
  w_{p\bar p} + w_{rr} = 0 , \label{lin8}
\end{equation}
while the f\/irst and second equations \eqref{redI_II} and their
complex conjugates take the form
\begin{gather}
  w_{pp}w_{\bar pr} - w_{p\bar p}w_{pr} - \lambda (w_{\bar pr}w_{zr} - w_{\bar
  pz}w_{rr}) = 0, \label{7I}
\\
 w_{p\bar p}(w_{p\bar p}w_{\bar zr} - w_{p\bar z}w_{\bar pr})
 + \lambda w_{\bar pr}(w_{\bar pr}w_{z\bar z} - w_{\bar pz}w_{\bar zr})
 \nonumber \\
\qquad{} + w_{\bar p\bar p}\big[w_{p\bar z}w_{pr} - w_{pp}w_{\bar zr}
 + \lambda (w_{zr}w_{\bar zr} - w_{z\bar z}w_{rr})\big] \nonumber \\
\qquad{} - w_{\bar p\bar z}\big[w_{p\bar p}w_{pr} - w_{pp}w_{\bar pr}
  + \lambda (w_{\bar pr}w_{zr} - w_{\bar pz}w_{rr})\big] = 0, \label{7II}
\end{gather}
respectively, together with their complex conjugates and equation~\eqref{bIIIred} reads
\begin{equation}
  w_{pp}w_{rr} - w_{pr}^2 - \lambda (w_{pr}w_{\bar pz} - w_{p\bar
  p}w_{zr}) = 0 \label{7bIII}
\end{equation}
after canceling nonzero factor $D$ due to condition \eqref{det}.
 Eliminating $w_{p\bar p}$ from \eqref{7I} with the aid of
 \eqref{lin8}, we obtain~\eqref{7I} in the form
 \begin{equation}\label{7_I}
  w_{\bar pr}(\lambda w_{zr} - w_{pp}) - w_{rr}(w_{pr} +
  \lambda w_{\bar pz}) = 0.
  \end{equation}
Similarly, using \eqref{lin8} in \eqref{7bIII}, we obtain
\begin{equation}
  w_{rr}(\lambda w_{zr} - w_{pp}) + w_{pr}(w_{pr} + \lambda w_{\bar pz}) = 0.
  \label{bilin}
\end{equation}
Determinant of the algebraic system of two latter equations is nonvanishing:
$w_{rr}^2 + w_{pr}w_{\bar pr} \ne 0$, since otherwise $w_{rr}
= 0$ and $w_{pr} = w_{\bar pr} = 0$ separately which contradicts
condition \eqref{det}.
Therefore, equations \eqref{7_I} and \eqref{bilin} become
\begin{equation}\label{linhom}
 w_{pp} - \lambda w_{zr} = 0,\qquad w_{pr} + \lambda w_{\bar pz} =
 0.
\end{equation}
Now we use linear equations~\eqref{lin8}, \eqref{linhom} and their complex
conjugates in equation~\eqref{7II} to obtain
\[
(w_{p\bar p} + w_{z\bar z})\big(w_{p\bar p}^2 + w_{p\bar z}w_{\bar pz}\big) = 0 ,
\]
which obviously implies
\begin{equation}
w_{p\bar p} + w_{z\bar z} = 0, \label{7_II}
\end{equation}
since otherwise $w_{p\bar p}^2 + w_{p\bar z}w_{\bar pz} = 0$ implies that
$w_{p\bar p} = 0$ and $w_{p\bar z} = w_{\bar pz} = 0$, which means
symmetry reduction with respect to physical variables that we wish
to avoid. Thus, we end up with the six linear equations~\eqref{lin8}, \eqref{linhom}, \eqref{7_II} and complex conjugates
of the two equations~\eqref{linhom}.
One can check that the complete system of six second-order linear equations
does not generate further second order integrability conditions.

We emphasize that these six linear equations were obtained by
symmetry reduction in the number of group parameters only, with
one remaining Legendre transformed group parame\-ter~$r$. No
reduction with respect to original ``physical'' variables has been
performed, so that we are able to obtain noninvariant solutions of
CMA by solving linear equations with constant coef\/f\/icients.
Noninvariant solutions of the general heavenly equation~\eqref{fer} can also be obtained by solving a linear system.

We note that Legendre transformation with respect to symmetry
group parameters was also used by M.~Dunajski and L.J.~Mason for
linearization of equations that determine invariant solutions of
the hierarchy of the second heavenly equation of Pleba\~nski~\cite{dunmas}.

\section{Transformation of K\"ahler metric}
 \label{sec-kahler}

Solutions $u=u(z^1,\bar{z}^1,z^2,\bar{z}^2)$ of the complex
Monge-Amp\`ere equation \eqref{cma} govern the K\"ahler metric
\cite{pleb}
\begin{equation}
 ds^2 = 2u_{1\bar 1}dz^1d\bar{z}^1 + 2u_{1\bar 2}dz^1d\bar{z}^2 + 2u_{2\bar
 1}dz^2d\bar{z}^1 + 2u_{2\bar 2}dz^2d\bar{z}^2 , \label{metr}
\end{equation}
which is an (anti-)self-dual Ricci-f\/lat metric and hence it
satisf\/ies vacuum Einstein equations with Euclidean signature.

To obtain a correct transformation of the metric \eqref{metr}, we
restrict transformation \eqref{Leg3} to physical variables only,
so that the Legendre transformation reads
\begin{gather}
 u = v - pv_p - \bar pv_{\bar p},\qquad z^1 = - v_p,\qquad
  \bar{z}^1 = - v_{\bar p},\qquad z^2 = z,\qquad \bar{z}^2 = \bar z, \nonumber  \\
  dz^1 = - (v_{pp}dp + v_{p\bar p}d\bar p + v_{pz}dz + v_{p\bar z}d\bar z), \nonumber \\
  d\bar{z}^1 = - (v_{\bar pp}dp + v_{\bar p\bar p}d\bar p + v_{\bar pz}dz
  + v_{\bar p\bar z}d\bar z), \label{Leg2}
\end{gather}
where $v$ is related to $w$ in \eqref{Leg3} by the one-dimensional Legendre transformation \eqref{leg1}.
Metric~\eqref{metr} after transformation~\eqref{Leg2} becomes{\samepage
\begin{gather}
 ds^2 = \frac{1}{\Delta}\left[v_{pp}(v_{\bar pp}d p + v_{\bar p
 z}dz)^2 + v_{\bar p\bar p}(v_{p\bar p}d\bar p + v_{p\bar z} d\bar
 z)^2\right] \nonumber \\
\phantom{ds^2 =}{} + \frac{\Delta_+}{v_{p\bar p}\Delta}(v_{\bar
 pp}d p + v_{\bar p z}dz)(v_{p\bar p}d\bar p + v_{p\bar z}d\bar z)
 + \frac{\Delta}{v_{p\bar p}}dzd\bar z , \label{tranmetr}
\end{gather}
where $\Delta = v_{pp}v_{\bar p\bar p} - v_{p\bar p}^2$, $\Delta_+
= v_{pp}v_{\bar p\bar p} + v_{p\bar p}^2$.}

We are able to construct many solutions $w=w(p,\bar p,z,\bar
z,r )$ of the system of six linear equations with constant
coef\/f\/icients: \eqref{lin8}, \eqref{7_II}, \eqref{linhom} and
complex conjugates of the two latter equations. In order to
reconstruct corresponding solutions $v(p,\bar p,z,\bar z,\rho)$
that govern metric~\eqref{tranmetr}, we will use the one-dimensional
Legendre transformation
\begin{equation}
  \rho = - w_r ,\qquad v = w - r w_r, \qquad r = v_\rho. \label{leg1}
\end{equation}
For given $w(p,\bar p,z,\bar z,r )$, equations \eqref{leg1}
determine the function $v(p,\bar p,z,\bar z,\rho)$ in a parametric
form where $r$ is the parameter. In examples of the next section,
we show how to obtain an explicit form of the potential $v$ of the
metric \eqref{tranmetr} by applying an inverse Legendre transformation
with respect to the group parameter to a solution $w$ of the linear system.

\section{Examples of exact solutions determining Ricci-f\/lat metrics\\ with no Killing vectors}
\label{sec-solutions}

Here we present some examples of non-invariant solutions of our equations
that determine Ricci-f\/lat metrics of the form \eqref{tranmetr} that do not
admit Killing vectors. For simplicity, we set $\lambda=1$
in equations \eqref{linhom}.

One general type of solutions of linear equations
\eqref{lin8}, \eqref{linhom}, \eqref{7_II} and complex conjugates
of equations \eqref{linhom} has the form
\begin{equation}\label{int}
    w = \int\limits_{\alpha_0}^{\alpha_1}F(\alpha,\xi_\alpha + i\eta_\alpha) d\,\alpha + \sum\limits_j F_j(\xi_{\alpha_j} + i\eta_{\alpha_j}) + \text{c.c.},
\end{equation}
where here and further $\text{c.c.}$ means complex conjugate of the previous term(s), $\alpha$ is real,
\begin{gather}
    \xi_\alpha = p + \bar p + i\alpha(\bar p - p),\qquad \eta_\alpha = \sqrt{\gamma}(r + \mu_\alpha),\nonumber\\
     \mu_\alpha = \frac{\alpha + i}{\alpha - i}z + \frac{\alpha - i}{\alpha + i} \bar z,\qquad \gamma = \alpha^2 + 1,\label{xieta}
\end{gather}
where bar means complex conjugate.

\subsection{Algebraic solution}
\label{sec-cubic}

Here we restrict ourselves to the case of the discrete spectrum in \eqref{int} and the cubic solution of the form
\begin{gather}
     w = a(\xi_\alpha + i\eta_\alpha)^3 + b(\xi_\beta + i\eta_\beta)^3 + \text{c.c.} \nonumber \\
\phantom{w}{}  = A\left[\xi_\alpha^3 - 3\gamma\xi_\alpha(r + \mu_\alpha)^2\right]
   + C\left[3\xi_\alpha^2(r + \mu_\alpha) - \gamma(r + \mu_\alpha)^3\right]\nonumber \\
\phantom{w=}{} + B\left[\xi_\beta^3 - 3\delta\xi_\beta(r + \mu_\beta)^2\right]
   + D\left[3\xi_\beta^2(r + \mu_\beta) - \delta(r + \mu_\beta)^3\right],
   \label{cubic}
\end{gather}
where $A$, $B$, $C$, $D$ are arbitrary real constants
\[A = a + \bar a,\qquad B = b + \bar b,\qquad C = i\sqrt{\gamma}(a - \bar a),\qquad D = i\sqrt{\delta}(b - \bar b),\]
$\xi_\beta$ and $\mu_\beta$ are obtained from $\xi_\alpha$ and $\mu_\alpha$
in \eqref{xieta} by replacing $\alpha$ by $\beta$, setting $\delta = \beta^2 + 1$ and $\eta_\beta = \sqrt{\delta}(r + \mu_\beta)$. Solution of the form \eqref{cubic} is the simplest nontrivial algebraic solution
in this class because a quadratic solution corresponds to constant metric coef\/f\/icients in \eqref{tranmetr} and hence to a vanishing connection and curvature tensor.

According to the Legendre transformation \eqref{leg1}, $r$ is determined in terms of $\rho$ and ``physical variables'' by the equation
$w_r + \rho = 0$. With $w$ determined by \eqref{cubic}, it takes the form of the quadratic equation for $r$
\begin{gather}
   kr^2 + 2lr + m = 0, \label{eqr}
\end{gather}
where
\begin{gather}
k = c\gamma + D\delta,\qquad l = \gamma(A\xi_\alpha + C\mu_\alpha) + \delta(B\xi_\beta + D\mu_\beta), \nonumber \\
  m = \gamma(2A\xi_\alpha\mu_\alpha + C\mu_\alpha^2) + \delta(2B\xi_\beta\mu_\beta + D\mu_\beta^2) - C\xi_\alpha^2 - D\xi_\beta^2 - \frac{\rho}{3}
   \label{klm}
\end{gather}
with the solution
\begin{equation}\label{root}
    r = \frac{- l \pm\sqrt{\Delta}}{k},\qquad\mbox{where}\qquad \Delta = l^2 - km.
\end{equation}
Next, we use \eqref{leg1} in the form $v = w + \rho r$ and replace $r$ by the expression \eqref{root} with the def\/initions \eqref{klm} to obtain
\begin{gather}\label{vsol}
  v =  \frac{1}{k^3}\big(P_3 \pm \sqrt{\Delta} P_2\big),
\end{gather}
where
\begin{gather*}
 P_2 = 6(A\gamma s_\alpha l_\alpha + B\delta s_\beta l_\beta) + C\left[3s_\alpha^2 - \gamma\left(\Delta + 3l_\alpha^2\right)\right]
             + D\left[3s_\beta^2 - \delta\left(\Delta + 3l_\beta^2\right)\right] + k^2\rho , \\
 P_3  =  A\left[s_\alpha^3 - 3\gamma s_\alpha\left(\Delta + l_\alpha^2\right)\right]
                + B\left[s_\beta^3 - 3\delta s_\beta\left(\Delta + l_\beta^2\right)\right] \nonumber \\
\phantom{P_3  =}{}  -   3Cl_\alpha\left[s_\alpha^2 - \gamma\left(\Delta + \frac{l_\alpha^2}{3}\right)\right]
                - 3Dl_\beta\left[s_\beta^2 - \delta\left(\Delta + \frac{l_\beta^2}{3}\right)\right] - k^2\rho l
\end{gather*}
and $s_\alpha = k\xi_\alpha$, $s_\beta = k\xi_\beta$, $l_\alpha = l - k\mu_\alpha$, $l_\beta = l - k\mu_\beta$. Here $P_2$ and $P_3$
are polynomials in $p$, $\bar p$, $z$, $\bar z$ of degrees $2$ and $3$ respectively.

By construction, the expression \eqref{vsol} for $v$ satisf\/ies the equation resulting after two dimensional Legendre transformation~\eqref{Leg2} of CMA equation \eqref{cma} with $\varepsilon = + 1$
\begin{equation}\label{veq}
    v_{p\bar z}v_{\bar pz} - v_{p\bar p}v_{z\bar z} + v_{pp}v_{\bar p\bar p} - v_{p\bar p}^2 = 0 ,
\end{equation}
which governs the Legendre-transformed metric \eqref{tranmetr}. Solutions \eqref{cubic} and \eqref{vsol} for $w$ and $v$ are obviously noninvariant,
since there is no symmetry reduction in the number of independent variables in these formulas. Therefore, solution \eqref{vsol}, being used in
the metric \eqref{tranmetr}, yields a~Ricci-f\/lat metric of Euclidean signature with no Killing vectors.

We can convert formula \eqref{vsol} for the solution to the polynomial form
\begin{equation}\label{polynom}
 k^6 v^2 - 2k^3 P_3 v + P_3^2 - \Delta P_2^2 = 0.
\end{equation}
By the change of variables $v = V^3$ and $\rho = \sigma^2$, solution \eqref{polynom} becomes
\begin{equation}\label{homogen}
    k^6 V^6 - 2k^3 P_3 V^3 + P_6 = 0 ,
\end{equation}
where $P_6 = P_3^2 - \Delta P_2^2$ is a homogeneous polynomial of degree $6$. Hence, our solution \eqref{homogen} is the set of roots of the homogeneous polynomial of degree $6$ in the six variables $V$, $p$, $\bar p$, $z$, $\bar z$, $\sigma$.
\begin{remark}
In our construction, we could skip the reality condition and replace the complex conjugate variables ${\bar z}^1$, ${\bar z}^2$ by two more independent complex
variables $\tilde{z}^1$, $\tilde{z}^2$. Then CMA equation \eqref{cma} is replaced by the f\/irst heavenly equation of Pleba\~nski \cite{pleb}
\begin{equation*}
   u_{1\tilde{1}}u_{2\tilde{2}} - u_{1\tilde{2}}u_{2\tilde{1}} = 1
\end{equation*}
and its Legendre transform has the form \eqref{veq} with $\bar p$, $\bar z$  replaced by $\tilde{p}$, $\tilde{z}$ respectively. Its algebraic solution is obtained from our real solution by the same replacements in formulas~\eqref{xieta} with $u$ and $v$ considered now as holomorphic functions of complex arguments and arbitrary real constants replaced by complex ones. Such a complex solution, being
the set of roots of the homogeneous polynomial of degree $6$ in the six complex variables $V$, $p$, $\bar p$, $z$, $\bar z$, $\sigma$,
determines a~four-dimensional compact manifold in a f\/ive-dimensional projective space $\mathbb{CP}^5$ with the local coordinates $(V/\sigma,p/\sigma,\bar p/\sigma,z/\sigma,\bar z/\sigma)$ \cite{GH}.

First example of an algebraic solution that governs anti-self-dual gravity was presented in our recent paper~\cite{mash}.

Ricci-f\/lat metrics in Euclidean signature with no Killing vectors on a compact manifold are interesting because they satisfy necessary existence conditions for the famous gravitational instanton $K3$~\cite{ahs}, whose explicit form is still unknown and presents an ``outstanding open problem in Riemannian geometry and the theory of gravitational instantons''~\cite{dunaj}.
\end{remark}

\subsection{Exponential solution}
\label{sec-exp}

Now we choose the exponential solution belonging to the discrete spectrum in \eqref{int}
\begin{equation}
   w = a e^{i(\xi_\alpha + i \eta_\alpha)/\sqrt{\gamma}} + \bar a e^{-i(\xi_\alpha - i \eta_\alpha)/\sqrt{\gamma}}
     + b e^{-i(\xi_\beta + i \eta_\beta)/\sqrt{\delta}} + \bar b e^{i(\xi_\beta - i \eta_\beta)/\sqrt{\delta}}
    \label{exp}
\end{equation}
with the same expressions for $\gamma$, $\delta$ and $\xi_{\alpha,\beta}$, $\eta_{\alpha,\beta}$ as before. After using these expressions in~\eqref{exp}
together with the def\/initions of arbitrary real constants
\begin{gather*}
    \cos{\theta}=\frac{1}{\sqrt{\gamma}},\qquad\sin{\theta}=\frac{\alpha}{\sqrt{\gamma}},\qquad\cos{\phi}=\frac{1}{\sqrt{\delta}},\qquad
   \sin{\phi}=\frac{\beta}{\sqrt{\delta}}, \\[2mm]
     A = a + \bar a,\qquad B = b + \bar b,\qquad C = i(a - \bar a),\qquad D = i(b - \bar b)
\end{gather*}
one obtains
\begin{equation}\label{w}
    w = e^{-r} G(p,\bar p,z,\bar z) + e^r H(p,\bar p,z,\bar z) ,
\end{equation}
where we have def\/ined
\begin{gather}
   G  = \exp{[\cos{(2\theta)}(z+\bar z) + i\sin{(2\theta)}(\bar z-z)]} \nonumber \\
 \phantom{G=}{}   \times  \{A\cos{[\cos{\theta}(p+\bar p) + i\sin{\theta}(\bar p-p)]} + C\sin{[\cos{\theta}(p+\bar p) + i\sin{\theta}(\bar p-p)]}\},
   \label{G}
\\
   H  =  \exp{[-\cos{(2\phi)}(z+\bar z) + i\sin{(2\phi)}(z-\bar z)]} \nonumber \\
\phantom{H  =}{} \times \{B\cos{[\cos{\phi}(p+\bar p) + i\sin{\phi}(\bar p-p)]} - D\sin{[\cos{\phi}(p+\bar p) + i\sin{\phi}(\bar p-p)]}\}.
   \label{H}
\end{gather}
The equation $w_r + \rho = 0$ following from~\eqref{leg1} now becomes the quadratic equation for the exponential~$e^r$:
$(e^r)^2 H + \rho e^r - G = 0$ with the solution for~$r$
\begin{equation*}
    r = \ln{\left\{\frac{- \rho \pm \sqrt{\rho^2 + 4GH}}{2H}\right\}}.
\end{equation*}
Using this expression for $r$ in equation \eqref{w} for $w$ and then in the equation $v = w + \rho r$ from~\eqref{leg1}, after elementary simplif\/ications
one obtains
\begin{equation}\label{v}
    v = \pm \sqrt{\rho^2 + 4GH} + \rho\ln{\left\{\frac{- \rho \pm \sqrt{\rho^2 + 4GH}}{2H}\right\}}
\end{equation}
with $G$ and $H$ def\/ined by~\eqref{G} and~\eqref{H}, respectively. By construction, the expression~\eqref{v} satisf\/ies the Legendre-transformed CMA
equation \eqref{veq} being a noninvariant solution of this equation, since there is obviously no symmetry reduction in the number of independent variables in this solution. Therefore, solution~\eqref{v}, being used in the metric~\eqref{tranmetr}, yields a~Ricci-f\/lat metric of Euclidean signature with no Killing vectors.

\section{Basic equations associated with the Husain equation}
 \label{sec-husain}

 Here we demonstrate that the same approach can be applied to the Husain equation \cite{husain}
\begin{equation}
  u_{ty}u_{xz} - u_{tz}u_{xy} + u_{tx} = 0 , \label{hus}
\end{equation}
which can be brought to a standard form by a simple change of variables. We shall brief\/ly
revise the derivation of the basic equations from Section~\ref{sec-eqs} in a form adapted to equation~\eqref{hus}.
If $\varphi$ is a symmetry characteristic of~\eqref{hus}, the symmetry condition reads
\begin{equation}
  u_{xz}\varphi_{ty} + u_{ty}\varphi_{xz} - u_{xy}\varphi_{tz} - u_{tz}\varphi_{xy} + \varphi_{tx} = 0.
  \label{symhus}
\end{equation}
Equation \eqref{symhus}, taken in an explicitly divergence form, implies the existence of the potential
$\psi$ connected with $\varphi$ by the relations
\begin{equation}
    \psi_t = \varphi_t + u_{ty}\varphi_z - u_{tz}\varphi_y ,\qquad
    \psi_x = u_{xy}\varphi_z - u_{xz}\varphi_y. \label{recurhus}
\end{equation}
The potential $\psi$ is a symmetry characteristic of the Husain equation~\eqref{hus} as soon as $\varphi$
is also a symmetry characteristic, so that \eqref{recurhus} determines a recursion relation for symmetries of~\eqref{hus}.
Using Lie equations~\eqref{Lie}, we replace symmetry characteristics  $\psi$, $\varphi$ in~\eqref{recurhus} by derivatives of $u$
with respect to the group parameters $\alpha$,  $\beta$ and $\gamma$, $\delta$, respectively:
\begin{gather}
  u_{\alpha t} = u_{\beta t} + u_{ty}u_{\beta z} - u_{tz}u_{\beta y},
  \label{1} \\
  u_{\alpha x} = u_{xy}u_{\beta z} - u_{xz}u_{\beta y},
  \label{2} \\
  u_{\gamma t} = u_{\delta t} + u_{ty}u_{\delta z} - u_{tz}u_{\delta y},
  \label{3} \\
  u_{\gamma x} = u_{xy}u_{\delta z} - u_{xz}u_{\delta y} .
  \label{4}
\end{gather}
Equations \eqref{1} and \eqref{3}, written in the form
\begin{gather}
  u_{\alpha t} = u_{\beta t} + (u_{ty}u_{z})_\beta - (u_{z}u_{\beta y})_t,
  \label{5} \\
  u_{\gamma t} = u_{\delta t} + (u_{ty}u_{z})_\delta - (u_{z}u_{\delta y})_t ,
  \label{6}
\end{gather}
imply the compatibility condition obtained by cross dif\/ferentiating \eqref{5} and \eqref{6}
with respect to $\delta$ and $\beta$, respectively, and integrating the result with respect to $t$
\begin{equation}
    u_{\alpha\delta} - u_{\beta\gamma} + u_{\beta y}u_{\delta z} - u_{\beta z}u_{\delta y} = 0.
    \label{compat}
\end{equation}
Here the ``constant'' of integration with respect to $t$ has been eliminated by a redef\/inition of~$u$.
Thus, we again have a system of six equations: \eqref{hus}, \eqref{1}--\eqref{4} and the integrability condition~\eqref{compat}.

Applying the same strategy as in Section~\ref{sec-symred}, we can perform symmetry reductions of this extended system
with respect to group parameters, which do not imply reductions in the number of original physical variables, and apply
an appropriate Legendre transform to obtain a system of easily solvable linear equations.

\section{Conclusion}

We have shown that noninvariant solutions of the complex
Monge-Amp\`ere equation can be obtained in principle by solving an
overdetermined system of six linear partial dif\/ferential equations
with constant coef\/f\/icients for one unknown depending on f\/ive
independent variables, that include a symmetry group parameter. We
considered in detail the complex Monge-Amp\`ere equation together with
recursion relations for partner symmetries. Using Lie equations,
we introduced explicitly four symmetry group parameters as
additional independent variables, which enabled us to derive one
integrability condition of the considered system. In this way, we
acquired four additional independent variables, so that we could make
symmetry reductions by eliminating three group parameters which
facilitated solution process of our equations but in no way meant
the reduction in the number of original ``physical'' variables.
Using Legendre transformation in three variables, including the
remaining group parameter, we arrived at linear partial
dif\/ferential equations with constant coef\/f\/icients which determine
symmetry group orbits of noninvariant solutions of the
Legendre-transformed CMA. These equations can be easily solved
and we have shown how to use these solutions for reconstructing explicitly
the potential $v$ that governs the Legendre transform of the
K\"ahler metric. We present algebraic and exponential examples of such
noninvariant solutions which determine Ricci-f\/lat
metrics with the Euclidean signature not admitting any Killing vectors.
We emphasize that our goal here was to obtain gravitational metrics with
the above properties for which it was suf\/f\/icient to obtain solutions of the
Legendre-transformed CMA equation that yield solutions of CMA in a
parametric form. To obtain an explicit form of the corresponding solutions
of the original CMA, we have to apply an inverse two-dimensional Legendre
transformation to our solutions for the potential $v$. We note that we
could consider the f\/irst heavenly equation of Pleba\~nski instead of the CMA
with our solutions still being valid if we consider the variables ${\bar z}^1$, ${\bar z}^2$
not as complex conjugates to~$z^1$,~$z^2$ but rather as additional independent complex variables.
We also note that in the same framework we may
choose for the reduction many other symmetries from our lists.
Solutions obtained with these alternative choices will be
published elsewhere. We have also brief\/ly outlined a similar approach
for the Husain equation.

\subsection*{Acknowledgements}

The authors thank M.~Dunajski for an interesting discussion about
the problem of performing the Legendre transformation of a metric
with respect to a symmetry group parameter.
The research of MBS was supported in part by the research grant
from Bogazici University Scientif\/ic Research Fund, research
project No.~07B301. AAM is grateful to Feza Gursey Institute for
the support of his stay at Istanbul where this work was f\/inalized.

\pdfbookmark[1]{References}{ref}
\LastPageEnding

\end{document}